\documentclass{ws-ijmpb}

\begin{document}

\markboth{Igor Trav\v enec}
{Metal - Insulator Transition in 3D Quantum Percolation}

%
%

\title{Metal - Insulator Transition in 3D Quantum Percolation}

\author{{Igor Trav\v enec}}

\address{Institute of Physics, Slovak Academy of Sciences, D\'ubravsk\'a cesta 9\\
SK - 84511 Bratislava, Slovakia.\footnote{ Institute of Physics, Slovak Academy of Sciences, D\'ubravsk\'a cesta 9, SK - 84511 Bratislava, Slovakia.}\\
fyzitrav@savba.sk}



\maketitle

\begin{history}
\received{Day Month Year}
\revised{Day Month Year}
\end{history}

\begin{abstract}
We present the metal - insulator transition study of a quantum site percolation model on simple cubic lattice. Transfer matrix method is used to calculate transport properties - Landauer conductance - for the binary distribution of energies. We calculate the mobility edge in disorder (ratio of insulating sites) - energy plane in detail and we find the extremal critical disorder somewhat closer to the classical percolation threshold, than formerly reported. We calculate the critical exponent $\nu$ along the mobility edge and find it constant and equal to the one of 3D Anderson model, confirming common universality class. Possible exception is the center of the conduction band, where either the single parameter scaling is not valid anymore, or finite size effects are immense. One of the reasons for such statement is the difference between results from arithmetic and geometric averaging of conductance at special energies. Only the geometric mean yields zero critical disorder in band center, which was theoretically predicted. 

\end{abstract}


\section{\label{sec:l1}Introduction\\}

Transport properties of disordered solids attracted lot of attention. The quantum percolation (QP) problem\cite{1,2} has been studied several decades and some questions are still open, compared to the more settled Anderson model\cite{3,4}. First of all, do both models belong to the same universality class? Positive answer means that the critical exponent $\nu$ of localization length should be the same, i. e. $\nu_A=1.57(2)$ in 3D and in presence of time reversal symmetry. The best data come from transfer matrix (TM) calculations of Anderson model\cite{5}. Another question is the validity range of single parameter scaling (SPS). It is possible, that SPS does not work in the middle of the conduction band for 3D QP, which is the case also for 1D Anderson model\cite{6,7}, but not for 3D Anderson model (at least numerically). Contrary to the still discussed 2D QP case\cite{8}, all authors agree that metal - insulator transition (MIT) is present in 3D, but they often disagree in details.

The mobility edge trajectory (MIT line in energy $E$ vs. disorder plane) in 3D is qualitatively clear and widely accepted \cite{9,10,11,12}, with possible jumps at special energies of incoming non-interacting transport careers ($E=0, \pm1, \pm\sqrt 2, ...)$\cite{13}. We aim to comment the nature of these jumps in more detail - for site percolation, but the bond percolation behaves similarly\cite{10}.

The common universality further requires constant $\nu$ along the mobility edge. The exponent should be the same both with respect to changing $E$ at fixed disorder and vice versa\cite{14,15}. We will check all these statements numerically. Another widely accepted fact is that $\nu$ should be identical both for site and bond percolation.

The values of the exponent $\nu$ given in literature differ heavily. Some methods seem to overestimate
the true value, if we accept $\nu_A$ as correct also for QP. Renorm-group calculations gave $\nu=2.2$ \cite{16} or $\nu=1.81(6)$ \cite{17}. Thouless conductivity\cite{18} yielded $\nu=1.95(12)$ on fcc lattice. On the other hand a special version of TM method with approximative reduction of matrices\cite{19} suggested $\nu=0.75(10)$ and cluster expansion with Pad\'e approximation\cite{20} yielded $\nu=0.38(7)$. The last value violates even the Chayes bound $\nu \ge 2/d$, where $d=3$ is dimension. It is known from the Anderson model, that level spacing methods tend to somewhat underestimate $\nu$ compared to the TM value; they suggest $\nu=1.35(10)$ \cite{21} and $\nu=1.45(7)$ \cite{22} for 3D QP. Thus we can say that the latter results confirm the common universality. The same is true of calculations using retarded Green's function method\cite{23}, with resulting $\nu=1.6(2)$ at $E=2$, but for small system sizes $L\le 8$ and rather spread data. Another paper of the same group\cite{12} reported lower $\nu=1.2(2)$. The main goal of this paper is to present another, TM based confirmation of the assumption, that Anderson model and QP belong to the same universality class.

\section{\label{sec:s2}Model and method}

Let us recall the tight-binding Hamiltonian:

\begin{equation}
\label{eq:1}
H = \sum_i \epsilon_i |i><i| + \sum_{<i,j>} {\rm t}_{ij} |i><j|
\end{equation}

\noindent where the first sum counts the $L^3$ lattice sites, the second sum goes over nearest neighbors and we set the hopping t$_{ij}=1$, thus fixing the energy scale. The length scale is set by choosing the lattice constant $a=1$. For a binary system the distribution of on-site energies $\epsilon_i$ has the form:

\begin{equation}
\label{eq:2}
P(\epsilon_i) = (1-q) \delta(\epsilon_i-\epsilon_A) + q \delta (\epsilon_i-\epsilon_B).
\end{equation}
We prefer to use $q$, the ratio of insulating atoms B, instead of often used $p=1-q$
because $q$ is the measure of disorder, induced among perfectly conducting A atoms.
The quantum percolation model can be got either by removing appropriate t$_{ij}\to 0$ in Eq. (\ref{eq:1}) or by sending $\epsilon_B \to \infty$. The latter case is more simple for TM method, as we need all t$_{ij}=1$ in the transfer direction and it suits better for site percolation. Thus we set the on-site energy for A-type atoms to $\epsilon_A = 0$ and the one for the randomly positioned B-type atoms to a large number\cite{9,10}, typically $\epsilon_B = 10^7$. The prize one has to pay in numerical approach are ill-defined matrices and a non-zero, but very small tunelling conductance through $B$ sites with still finite barrier - though the latter disadvantage shows up
only in deep insulators \cite{8}, behind the scope of this paper.

For Anderson model the $P(\epsilon_i)$ in Eq. (\ref{eq:2}) is continuous, e. g. box distribution from an interval of the width $W$ (disorder):  $\epsilon_i \in [ - W/2, W/2 ]$. It should be further mentioned, that both models are fully symmetric with respect to $E \rightarrow -E$, thus we will restrict our investigation to one (left) half of the  band:\ $-6 < E \le 0$. 

The classical site percolation threshold for simple cubic lattice is $q_C=0.6884$ and it is known, that the quantum critical value $q_Q < q_C$ for the whole mobility edge. The quantum wave requires a better highway than the narrow path of the percolation cluster backbone. Up to now the maximum calculated value\cite{10,23} was $q_Q \approx 0.56(1)$, but we will show that the difference from $q_C$ is smaller.

Our numerical calculations are based on the usual transfer matrix method, described elsewhere\cite{4}. The conductance $g$ in units $e^2 / h$ is given by the Landauer formula $g= 2\ {\rm Tr}\ t^+t$. Here $t$ is the $L^2\times L^2$ transmission matrix, specifying the transmitted (not reflected) part of $L^2$ planar waves, entering the sample on the left and leaving it on the right. $t$ is extracted from a $2L^2\times 2L^2$ transfer matrix. It was mentioned above, that we have to do with ill-defined matrices, even more than within Anderson model. In transfer matrix approach we deal rather with $t^{-1}$ matrices and the precision of their small eigenvalues gets quickly lost - but we need them accurate, as they become important after recovering the $t$ matrix. An efficient way how to treat this problem was described in Refs.~\refcite{24} and \refcite{4}; we have to perform the renormalization procedure at least after each two TM multiplications, but then we are able to include larger samples up to $L=20$. This is useful, as finite size effects turn out larger than for Anderson model. The most numerically sensitive region is around the band center, i. e. for small $|E|$, where we have to perform the procedure after each single TM multiplication\cite{8}.

Once we have the matrix $t^\dag t$, we can calculate also its eigenvalues $\tau_i$, satisfying

\begin{equation}
g=2\ \textrm{Tr}\ t^\dag t=2\sum_i^N \tau_i= 2\sum_i^N{\frac{1}{\cosh^2 x_i/2}}
\end{equation}
where the positive quantities $x_i$ play similar role for cubic samples, as Lyapunov exponents do in quasi-1D geometry. The smallest one is labeled $x_1$.

In localized samples, the Landauer conductance decays quickly with sample size, $g\sim \exp(-L/\xi)$. The correlation length $\xi(q,E)$ diverges at critical point $(E_Q, p_Q)$ as $\xi \sim (q-q_Q)^{-\nu}$ for $E_Q$ fixed and vice versa. The SPS assumption implies one combined variable from $(q,L)$ or $(E,L)$ and in the vicinity of the MIT point we expect $g(q,L)\sim g(L/\xi(q))\sim g(L\ (q-q_Q)^{\nu})$. If we perform Taylor expansion in $(q-q_Q)$ variable and add some finite-size correction, we get

\begin{equation}
\label{exp}
g=g_Q+A_1 (q-q_Q)L^{1/\nu} + A_y L^{-y}+O\Bigl(\bigl[(q-q_Q)L^{1/\nu}\bigr]^2\Bigr),
\end{equation}
where $y$ is an irrelevant exponent and its term becomes negligible for large $L$. Analogous relations can be expected for $q \leftrightarrow E$, i. e. with the single variable $(E-E_Q)L^{1/\nu}$ and $q_Q$ fixed. The conductance $g$ will be mostly an ensemble average $\langle g \rangle$, taken over many samples, typically 20.000. It is known\cite{25} that many other quantities would fit Eq. (\ref{exp}), e. g. the geometric mean $\exp(\langle \ln g \rangle)$, $1/\langle 1/g \rangle$, percentiles of the distribution of conductances $P(g)$, $\langle x_1 \rangle$, current moments\cite{26} $\langle \sum_i \tau_i^n\rangle$, inverse participation ratios\cite{15}, etc\cite{13}. Nevertheless, quantities like $\langle \ln g \rangle$ or $\langle x_1 \rangle$ are more sensitive to numerical instabilities, especially for larger $p$ and $L$ or very small $|E|$.

\section{Numerical results}


It was shown\cite{16}, that the density of states (DOS) has sharp peaks at special energies, $|E|=0,1,\sqrt 2, (3\pm\sqrt 5)/2, ...$ These eigenvalues belong to submatrices of the Hamiltonian Eq. (\ref{eq:1}), created by small islands of $A$ sites completely surrounded by $B$ sites, i. e. isolated from the rest. It was found, that $g$ can quickly fall down at these energies and they can influence the shape of mobility edge.

\begin{figure}[bt]
\vspace*{17pt}
\centerline{\psfig{file=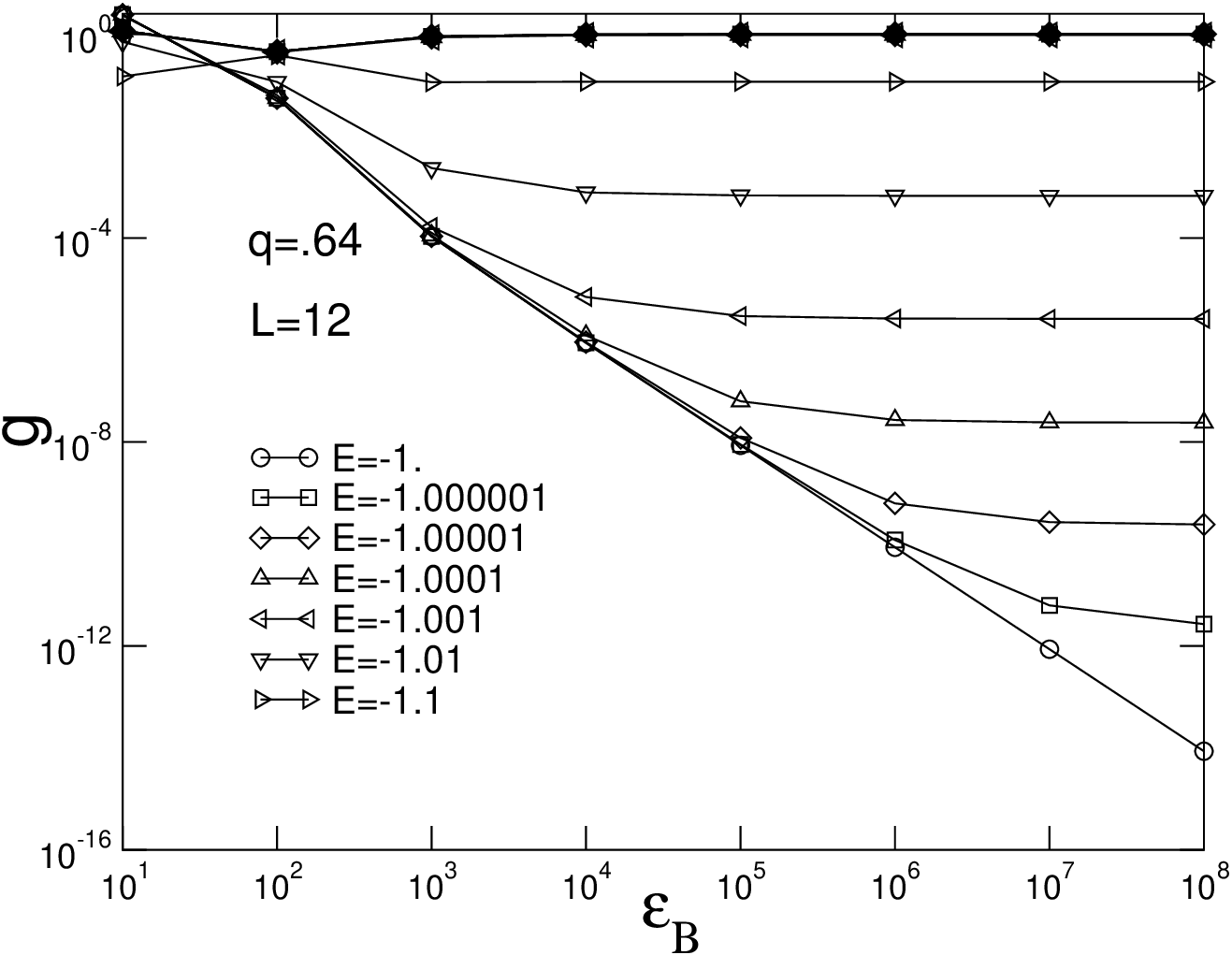,width=3.65in}}
\vspace*{8pt}
\caption{Conductance $g(\epsilon_B)$ near special point $E=-1$. Two samples: the first, delocalized one - overlaying full symbols above; another one - empty symbols reaching from slightly to deeply localized case.}
\label{fig:f1}
\end{figure}

Before going to main results, let us check our numerics in the vicinity of the special point $E=-1$. Conductance $g$ should be $\epsilon_B$ independent for large enough $\epsilon_B$. We can see in Fig. \ref{fig:f1}, that for delocalized sample with $g \ge 1$, common upper line, this is true for $\epsilon_B \ge 10^4$ even at $E=-1$. For a localized sample, all lower lines, the value depends strongly on the energy $E$ close to the special $E=-1$, but still a plateau develops for larger $\epsilon_B$. Exception is the special point itself, where $g\propto \epsilon_B^{-2}$, in other words $g \to 0$ for $\epsilon_B \to \infty$. The same $g(\epsilon_B)$ dependence is found for a test sample with one perpendicular square barrier of $B$ sites completely stopping the transport through otherwise perfect cube of $A$ sites; thus this dependence is an unusual form of tunelling. The picture is qualitatively the same at $E=0$, just for much lower disorder $q$. It is worth mention, that in the latter case we surely do not have to do with simple quantum tunelling through closed finite samples\cite{26} (closed in the sense of classical percolation, regardless of $E$), as we chose samples with $q=0.37$, far bellow $q_C$. Alternatively formulated, the same samples become metallic for larger $|E|$. One can also see that the geometric mean of conductance tends to zero though the arithmetic $\langle g \rangle$ remains positive at special points. We chose also $E=-\sqrt{2}$ and after some effort we found a localized sample of the type as in Fig. \ref{fig:f1} bottom lines, but the ratio of such specially localized samples becomes much lower than for $E=0,\pm 1$. 

The possible deep localization at special energies was explained in several ways. In Ref. \refcite{1} it was supposed, that the clusters of insulating atoms around conducting islands could have extremely large reflectivity at these energies. Another explanation\cite{13} argued that localization on dead ends of the spanning cluster is strong at special energies. Let us add yet another possibility. The dispersion law gives

\begin{equation}
\label{eq:5}
E = -2 {\rm t} \cos k_z a -2 {\rm t} \cos {\frac{\pi n a}{L+1}} -2 {\rm t} \cos {\frac{\pi m a}{L+1}}, \ \ \ 0 \le k_n \le \pi/a; \ n, m = 1, ..., L
\end{equation}
where $k_z$ is the wavenumber in transfer direction $z$ and the wavenumbers in perpendicular directions $x, y$ are quantized by hardwall boundary conditions. For special values of energies the whole vector ($k_x, k_y, k_z$) can point in (body) diagonal direction and the planar wave is then fully reflected by insulating site, siting exactly in this direction. If all planar waves are fully reflected, the sample becomes deeply localized\cite{27}.

Summarizing, with $\epsilon_B \simeq 10^7$, we always get correct $g$ for delocalized samples and also for medium localization. Strongly localized samples require special attention. 
\begin{figure}[htb]
\vspace{5pt}
\includegraphics[width=.43\textwidth]{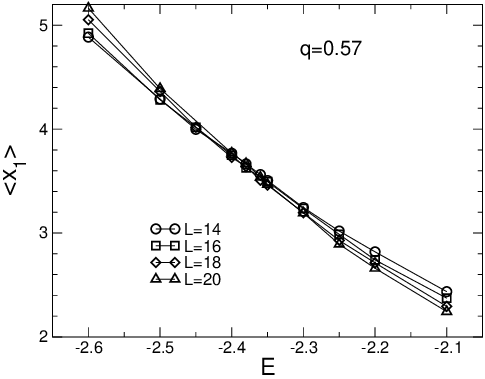}~(a)
\hfil
\hspace{2pt}
\includegraphics[width=.43\textwidth]{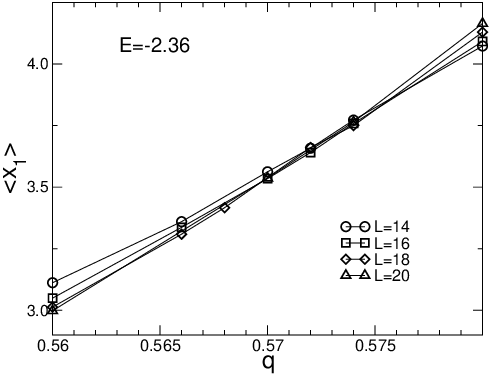}~(b)
\caption{(a) $\langle x_1\rangle$ vs. energy and (b) $\langle x_1\rangle$ vs. disorder, both around the same MIT point $(q_Q=0.57, E_Q=-2.36)$.}
\label{fig:f2}
\end{figure}

Now let us find one critical point in detail. For $q<q_Q$ most samples are metallic and $\langle g \rangle \propto L^{d-2}\propto L$, for $q>q_Q$ most samples are localized and $\langle g \rangle \propto \exp(-L/\xi)$. At MIT, $g$ becomes $L$ independent. We chose the value $q_Q=0.57$, and found $E_Q=-2.36$ from the crossing point in Fig. \ref{fig:f2}a) and \ref{fig:f2}b). As $q_Q$ is far enough from $q_C$, and $E_Q$ is not close to any important special value (in the sense of relevant peaks in DOS), we can use even the $\langle x_1\rangle$ as order parameter. The data were fitted according to Eq. (\ref{exp}) and the resulting exponent is $\nu=1.6(1)$ for the $(q-q_Q)$ dependence and $\nu=1.5(1)$ for the $(E-E_Q)$ one, both in perfect accordance with $\nu_A$. Of course, taking $\langle g\rangle$ instead of $\langle x_1\rangle$ yields the same critical point and $\nu$.

\bigskip
\bigskip
\medskip
\begin{figure}[bt]
\vspace*{17pt}
\centerline{\psfig{file=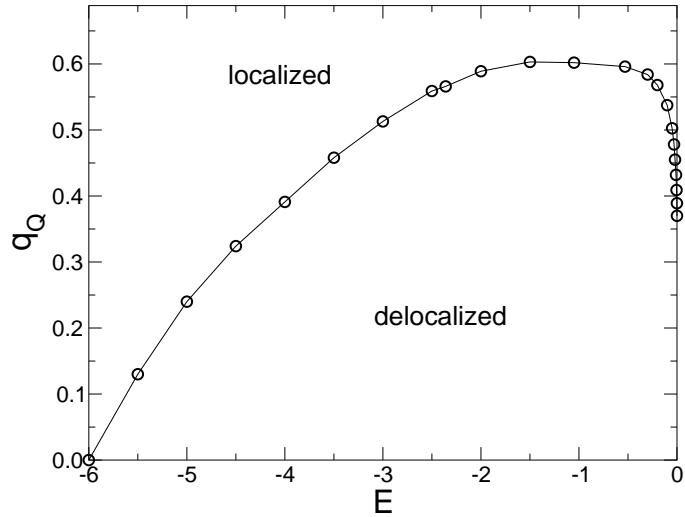,width=3.55in}}
\vspace*{6pt}
\caption{The mobility edge, or MIT line.}
\label{fig:f3}
\end{figure}

\begin{figure}[htb]
\vspace{5pt}
\includegraphics[width=.43\textwidth]{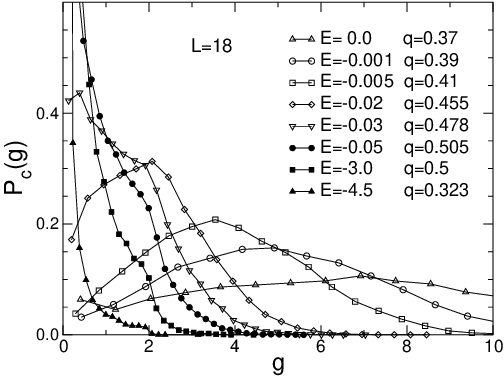}~(a)
\hfil
\hspace{2pt}
\includegraphics[width=.43\textwidth]{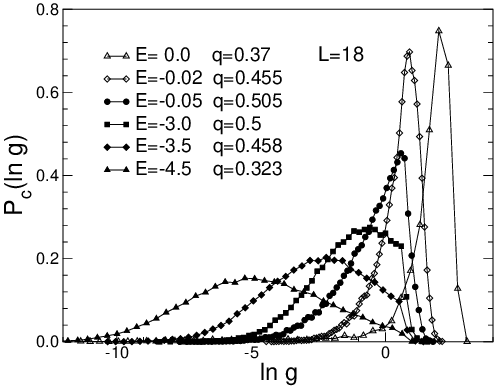}~(b)
\caption{Critical distributions of $g$ (a) and $\ln g$ (b) along the MIT line.}
\label{fig:f6}
\end{figure}

Having performed these calculations for almost two dozens of fixed energies, typically changing only $q$, we can plot the mobility edge in Fig. \ref{fig:f3}. Some pairs of numerical values can be found in Fig. \ref{fig:f6}. The point $(-6,0)$ was added artificially - there are no open channels there. It was already mentioned above that the situation is mirror symmetric in $E\rightarrow -E$, we do not plot the obvious interval $0 < E \le 6$. The upper border of the graph was set to the classical $q_C$. The maximum value of our MIT line is cca $q=0.603(2)$, greater than $0.56(1)$ value reported before\cite{10,12}, but with smaller samples $L\le 9$ and $L\le 8$, resp. 

\begin{figure}[htb]
\vspace{5pt}
\includegraphics[width=.43\textwidth]{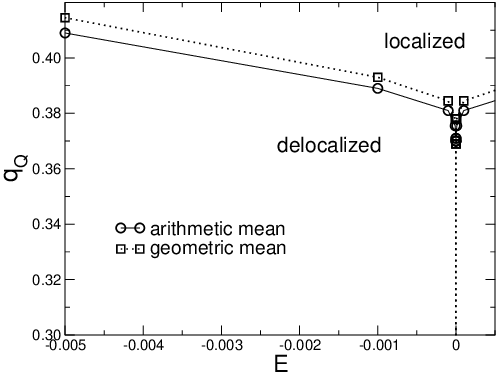}~(a)
\hfil
\hspace{2pt}
\includegraphics[width=.43\textwidth]{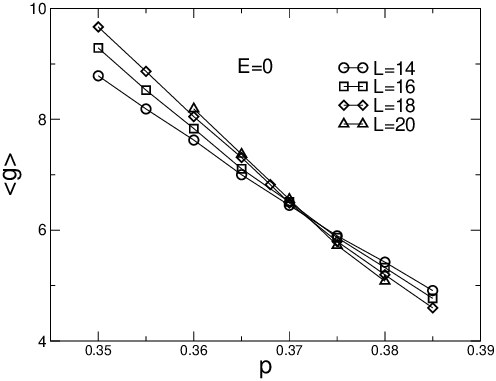}~(b)
\caption{(a) Detailed view of the mobility edge at band center, (b) $L$ - independent crossing point of arithmetic mean $\langle g(p)\rangle$ at band center.}
\label{fig:f4}
\end{figure}


Now we will compare our results to those of Ref. \refcite{13}. They introduced the local DOS for each lattice site and then performed two averages at once (in two manners). The first one was ensemble average as we do, the second one was averaging with respect to all sites. The two manners were geometric and arithmetic mean. At last they calculate the ratio of geometric/arithmetic mean and plot lines, where this order parameter is constant. The line entering the point $(-6,0)$ compares very well with our mobility edge in Fig. \ref{fig:f3}, with their minimum at cca $p=0.4$, i. e. our maximum at $q=0.6$ and a rather wide plateau around this maximum. The main difference is a discrete set of energies, where their order parameter and thus also the critical line drops to zero. But the only cases, where they could show this explicitly where integers $E=0, \pm1$. With irrational values like $E=\pm\sqrt{2}$ the localized case is hard to find numerically, as its region is extremely narrow in energies and also the corresponding peaks in DOS are much smaller\cite{16} than those at three integer $E$. Hence we plot the detailed view of the band center in Fig. \ref{fig:f4}(a), where both calculations of critical $q_Q$ with the use of geometric and arithmetic mean of conductance $g$ are presented - of course only ensemble averaged. The critical disorder $q_Q$ from calculations with geometric mean of $g$ is systematically somewhat larger outside the band center, but only by several percent - beware that the $q$ axis is not drawn to zero. For very small $|E|=10^{-6},\ 10^{-7}$ the $q_Q$ from geometric mean already becomes slightly smaller than that from arithmetic mean and directly at $E=0$ it drops steeply to zero. It should be stressed out, that the decay of MIT line at band center down to cca $q_Q=0.37$ is independent of the type of averaging, it happens within a clearly non-zero interval of energies and it should be distinguished from extremely narrow drops of MIT line to zero at special energies, present only with geometric averaging. Similar relatively broad interval of energies, where the mobility edge should decay, was reported\cite{10} around $E=1$. With arithmetic mean and $\epsilon_B=10^7$, we do not see it at all. We should also say that the standard mean DOS, calculated up to $L=16$, has clearly non-zero value for $q=0.37$ in the vicinity of the band center, apart from high central peak at $E=0$. According to our calculations, only at cca $q=0.44$ the DOS drops to zero for small $|E|$ and a gap opens for even larger $q$.

The exponent $\nu$ remains constant along the whole mobility edge line, just the error bars rise as we approach the borders of the band $|E|=6$ and $E=0$. But even for $E=-0.001$ and $q_Q=0.39$ it was still $\nu=1.5(3)$. Important exception is the point $E=0$ itself. Only with geometric mean, it has $q_Q=0$, as suggested already in Ref.~\refcite{1}. One possible explanation is that single parameter scaling (SPS) does not work here anymore. The argument is, that geometric mean of $g$ differs substantially from the arithmetic one. Furthermore, if we insists on SPS and arithmetic $\langle g \rangle$ as order parameter, standard analysis of data from Fig. \ref{fig:f4}(b) according to Eq. (\ref{exp}) yields $q_Q=0.37$ at $E=0$ and $\nu \approx 1$ with $(q-q_Q)$ variable. We tried also to calculate  $\nu$ with $(E-0)$ variable, but the $g(E)$ dependences changed their slope below $|E| < 10^7$ and the analysis was too unstable. It is improbable, that common universality would be broken at one spectral point, the $\nu$ from Fig. \ref{fig:f4}(b) data smaller than $\nu_A$ could rather mean, that SPS is broken at $E=0$. But one can also argue, that finite size effects are so strong here, that even sample sizes up to $L=20$ are insufficient. This would be exceptional, though not completely excluded. In Ref. \refcite{15} the authors had similar problems with 3D Anderson model at the band edge (close to $E=6$) and explained them by insufficient available sample sizes. Summarizing, we always get a well defined crossing point of mobility edge, where data become $L$-independent as they should at criticality, even for $E=0$, see Fig. \ref{fig:f4}(b). Just the slopes of the crossing lines behave non-universally in band center, at least for available system sizes.

\bigskip
\bigskip
\medskip
\begin{figure}[bt]
\vspace*{16pt}
\centerline{\psfig{file=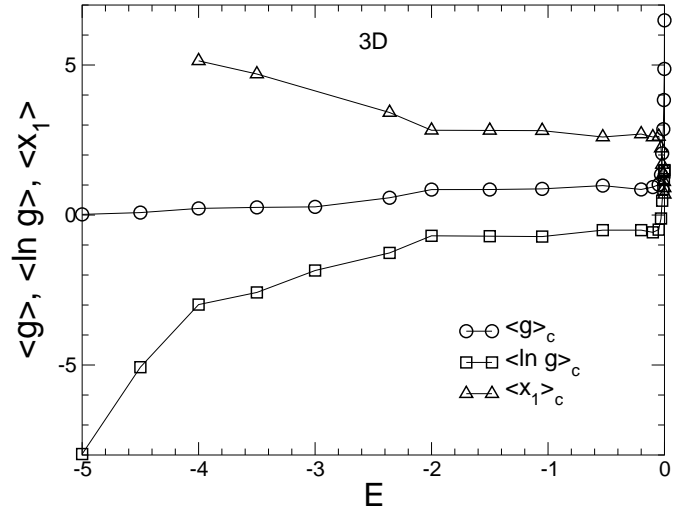,width=3.5in}}
\vspace*{5pt}
\caption{Critical values of $\langle g\rangle$, $\langle \ln g\rangle$ and $\langle x_1\rangle$ along the mobility edge.
}
\label{fig:f5}
\end{figure}

Let us present mean critical values of several quantities, namely $\langle g\rangle_C$, $\langle \ln g\rangle_C$ and $\langle x_1\rangle_C$. For $g < 2$ we can roughly say that it is given mainly by the largest eigenvalue $g \approx 2\tau_1 = 2 \cosh^{-2}(x_1/2)$ and after simple algebra $\ln g \approx 3\ln 2 - x_1$, which is well fulfilled in Fig. \ref{fig:f5}, of course only outside the band center. For small $|E|$ we get $g_C>2$ as a sum of several important $\tau_i$.

It should be emphasized, that $\langle g\rangle_C$ remains finite, but $\langle \ln g\rangle_C$ and $\langle x_1\rangle_C$ diverge, as we approach $E = 0$, see comments to Fig. \ref{fig:f1}. The divergence is given by a minority of fully localized samples with $g \to 0$ for $\epsilon_B \to \infty$, which do not really influence the arithmetically averaged $g^{max}_C\approx 6.5$.

Further we present several critical conductance distributions $P_C(g)$ and $P_C(\ln g)$ along the mobility edge.  
The distribution at $E=0$ and $p=0.37$ was added for comparison, too. 

We do not plot critical distributions for $-0.05 > E > -3$ which are mutually almost overlaying, because of similar $q_Q$. Further we can see typical non-analyticity\cite{4} at $g=2$ or $\ln g = \ln 2$, if $P(g)$ or $P(\ln g)$, resp., is already decaying there. The $P_C(\ln g)$ at $E=-3$ and those with maximal $q_Q$ look similar to $P_C(\ln g)$ of 3D Anderson model, but they are not identical for any $(E,q)$ - the SPS works separately inside the models, not universally. This is known also for anisotropic Anderson models and those in other dimension than $d=3$. Even for the bifractal\cite{27,4} with the lowest $d = 2.226$ and $g_C \approx 7$ the $P_C(g)$ looked more Gaussian than those in Fig.\ref{fig:f6}(a) on the right. This is connected with far larger ${\rm var}\ g = \langle g^2\rangle-\langle g\rangle^2$ of 3D QP compared to low - dimensional Anderson model at criticality or to universal conductance fluctuations in metallic regime. 

Let us shortly discuss the metallic regime. Away from band center, say with $E=-1.05$, we found typical metallic behavior for medium disorder $0.2 \le q \le 0.5$, i. e. $P(g)$ practically Gaussian, $\langle g\rangle \propto L$ and var $g$ constant ($L$ and $q$ independent), with values cca 10 percent above the theoretically predicted universal conductance fluctuations\cite{4}. Even smaller $q$ brings traces of ballistic behavior, var\ $g$ becomes system-size dependent (rising with $L$). Contrary to these standard properties, at the band center $E=0$ we did not find typical metal for any $q$ and available $L$. We still have $\langle g\rangle \propto L$, but now ${\rm var}\  g\propto L$ instead of reaching the universal constant and cca one order of magnitude greater - rather like the mentioned almost-ballistic behavior (perfect ballistics would have $\langle g\rangle \propto L^2$). Also $P(g)$ is somewhat deformed, not Gaussian. The transition from this behavior to standard metal at larger $|E|$ is continuous, var $g$ decays with rising $|E|$. If this is only finite size effect, than it is again immense at band center. 

\section{Conclusions}

We already stated recently\cite{8}, that 2D quantum percolation problem and 2D Anderson model belong to the same universality class, meaning that $d=2$ is lower critical dimension for both models and there is no MIT in time reversal symmetry case. Here we add the confirmation of common universality in 3D, implying identical critical exponent $\nu$, which we have shown within numerical accuracy along the mobility edge, with possible exception of the conduction band center $E=0$. Another new result confirms equal $\nu$ when cutting the mobility edge (MIT line) either in disorder or in energy direction. All these properties result from single parameter scaling hypothesis and they get lost at band center, where either the SPS does not work anymore, or we face immense finite-size effects, that cannot be overcome with system sizes up to $L=20$. This statement is supported also by loosing many typical properties of metallic samples at band center. Concerning the mobility edge, we found the maximum critical disorder $q_Q$ slightly above 0.6, which is more than $q_Q \approx 0.56$ reported before\cite{10,23}. The shape of the mobility edge at band center depends on the choice of order parameter. The arithmetically averaged $\langle g\rangle$ (commonly preferred) gives clearly non-zero $q_Q \approx 0.37$ at $E=0$, whereas the geometrical mean gives $q_Q=0$ in accordance with other numerics\cite{13} and theory\cite{1}. We present a new alternative explanation of extremely low conductance for a minority of samples, which push the geometric mean to zero. It is based on the fact, that for special energies the wavevectors of incoming planar waves point in body diagonal directions and the waves can face exactly the insulating sites, followed by full reflection.

The rational special energies $E=0, \pm1$, have clearly different arithmetic and geometric means and also the resulting $q_Q$. The irrational ones, e. g. $E=\pm \sqrt{2}$, seem numerically almost insensitive of the type of averaging, as the samples with nearly zero conductance become very rare. But the rather steep descent of mobility edge at band center takes place at non-zero interval of energies with SPS still working for both types of averaging and it should be distinguished from the extremely narrow regions at all special energies, apparent only for geometric mean of conductances.

\section*{Acknowledgements}

This work was supported by Grant VEGA Nr. 2/6069/26.

\end{document}